\documentclass[aps,pre,reprint,superscriptaddress]{revtex4-2}
\usepackage{amsmath,amssymb,amsfonts,graphicx,color,soul,ulem}
\usepackage[english]{babel}

\usepackage{kotex}

\begin{document}


\title{Searching for a partially absorbing target by a run-and-tumble particle in a confined space}


\author{Euijin Jeon}
\affiliation{Department of Physics, Technion–Israel Institute of Technology, Haifa 3200003, Israel}
\author{Byeongguk Go}
\author{Yong Woon Kim}
\affiliation{Department of Physics, Korea Advanced Institute of Science and Technology, Daejeon 34141, Korea}


\date{\today}

\begin{abstract}
A random search of a partially absorbing target by a run-and-tumble particle in a confined one-dimensional space is investigated.
We analytically obtain the mean searching time,
which shows a non-monotonic behavior as a function of the self-propulsion speed of the active particle, indicating the existence of an optimal speed,
when the absorption strength of the target is finite.
In the limit of large and small absorption strengths, respectively, asymptotes of the mean searching time and the optimal speed are found. 
We also demonstrate that the first-passage problem of a diffusive run-and-tumble particle in high dimensions can be mapped into a one-dimensional problem
with a partially absorbing target.
Finally, as a practical application exploiting the existence of the optimal speed, we propose a filtering device to extract active particles with a desired speed
and evaluate how the resolution of the filtering device depends on the absorption strength.
\end{abstract}


\maketitle

\section{Introduction}
How long will it take for a blind searcher to find a target in a confined space?
This is a central question in the random target search or the first-passage process~\cite{Redner2001}, which plays a pivotal role in understanding a variety of phenomena ranging from diffusion-limited reactions~\cite{Smoluchowski1917, Kramers1940, Tachiya, Krapivsky2014, Benichou2010, Metzler2019, Isaacson2009}, predator foragings~\cite{redner_capture_1999} to intracellular protein transports~\cite{sheinman_classes_2012}.
There have been a number of studies on the random target search, for example, with a focus on the optimal search strategy~\cite{o1990search,raposo_dynamical_2003,PhysRevLett.94.198101,levernier2020inverse,meyer2021optimal}, the influence of space topography~\cite{noh_random_2004,volpe2017topography}, the dependence on the number or the initial distribution of searchers~\cite{ro2017parallel,lawley2020probabilistic,snider_optimal_2011, Kim2022}, and the effect of interactions between searchers~\cite{drager_mean_2000,agranov_survival_2016, Kim2021, Kim2023}.

Active matter are nonequilibrium systems that are driven by consuming energy, and they have received a lot of attention over the past decades because they exhibit distinct characteristics that distinguish them from passive particles, not only in the collective behaviors~\cite{toner_flocks_1998, romanczuk2012active, woodhouse2012spontaneous,cates_motility-induced_2015,bechinger_active_2016} but also at the individual particle level~\cite{szamel2014self, Solon2015NatPhys, Angelani_2017, o2021time}.
Two popular models for active matter are the run-and-tumble particle (RTP) and the active Brownian particle (ABP).
RTP switches between the run phase, performing a straight motion in a certain direction, and the tumbling phase, randomly reorienting the propulsion direction.
ABP is subject to thermal noises and changes the self-propelling direction by the rotational diffusion.
For active particles, the persistence length $\xi$ can be defined as a length scale over which the orientation of the trajectory is maintained,
which is given as $\xi=v/\lambda$ for RTP with the propulsion speed $v$ and the tumbling rate $\lambda$, and $\xi=v/D_\theta$ for ABP with the rotational diffusion coefficient $D_\theta$.

The random target search by active particles has been recently studied in the context of finding optimal strategy.
In a two or higher dimensional space, it was shown that the optimal search strategy to minimize the traveling distance is to have a finite persistence length~\cite{PhysRevLett.108.088103}:
Large persistence length is advantageous because it prevents the redundant exploration of the same region, while too large persistence length is disadvantageous because there is a risk that the searcher is trapped in a long unsuccessful excursion.
The optimal persistence length has been first obtained analytically for a two-dimensional discrete lattice system with periodic boundary condition\cite{PhysRevLett.108.088103}, and later numerically simulated for a continuum space with an active searcher modeled by RTP\cite{rupprecht2016optimal}  or ABP\cite{doi:10.1063/1.4952423}.

Most of the previous studies on the random target search by an active particle considered a ideal, perfect target described by a fully absorbing boundary so that the searcher finds the target with certainty at the first encounter.
However, in reality, there exists a partially absorbing target and the searcher finds the target only probabilistically when it reaches the target boundary~\cite{Guerin2021}.
In particular, this is the case for diffusion-limited reactions where a pair of molecules react with a finite reaction energy when coming to the distance of interaction ranges.
Although there have been several studies on the random search for a partially absorbing target by a passive particle
\cite{  Grebenkov2010,Isaacson2016,Guerin2021}, the same problem by an active particle is still unexplored.

In this work, we address this problem by considering a RTP in a confined one-dimensional space.
Incorporating the presence of a partially absorbing target as a sink term in the master equation, we obtain the analytic expression of the mean searching time
as a function of a propulsion speed and a reactivity of the target and show that for a finite reactivity, there exists an optimal speed of RTP to minimize the search time.
Asymptotic behaviors of the mean searching time and the optimal speed are derived in the limit of small and large reactivities, respectively.
Using the optimal speed, we also propose a filtering device to extract active particles of a certain speed from a mixture of particles with different velocities
and evaluate the resolution of the filtering device to be achieved in terms of the system parameters.

This paper is organized as follows.
In Sec.\ref{sec2}, the mean searching time of a RTP in the presence of reactive target is evaluated analytically.
In Sec.\ref{sec3}, the asymptotic behaviors of the mean searching time and the optimal speed are examined.
Section \ref{sec4} is devoted to the applications of the present study to high dimensional systems and a filtering device of active particles. 
The conclusion is given in Sec.\ref{sec5}.

\section{Run-and-tumble particle with a partially absorbing target}\label{sec2}

We consider a one-dimensional run-and-tumble particle (RTP) confined by hard-walls at $x=\pm L$.
RTP undergoes a straight motion with a constant speed $v$ ({\it run}) until the direction of the motion is randomly reversed ({\it tumble}).
The equation of motion is simply described by
\begin{equation}
\dot{x}=\sigma v
\end{equation}
where $\sigma = \pm 1$ is the direction of the motion, which flips its sign in the Poisson process with a rate $\lambda$.
Instead of perfect absorption at the target location, we consider a partially absorbing target (also called a reactive target), i.e., upon encountering, the target is not recognized with probability one.
For diffusing particles, a partially adsorbing boundary is usually described by the Robin (also known as reactive) boundary condition~\cite{Collins1949, Isaacson2016},
which can be derived from a diffusion equation with an effective sink term~\cite{Isaacson2016, Whitehouse2013}.
In the same way, we incorporate the presence of a partially absorbing target in the dynamical equation (telegrapher's equation) for the probability distribution function
$p_{\sigma}(x, t)$ of RTP to be at position $x$ at time $t$ with direction $\sigma$, written as
\begin{equation}\label{master}
\partial_t p_\sigma (x, t)=-\sigma v\partial_x p_{\sigma}-\lambda \left(p_\sigma-p_{-\sigma}\right) - k \delta (x) p_{\sigma} (x, t)
\end{equation}
where $\partial_t=\partial/\partial t$ and $\partial_x=\partial/\partial x$ are the temporal and positional derivatives.
The partially absorbing target at the origin is represented by a sink term with $k$ being the absorption strength (reactivity) of the target.
In order to obtain the boundary condition from the sink term, we divide both sides with $p_{\sigma}$, leading to
\begin{equation}
\partial_t \ln p_\sigma =-\sigma v\partial_x \ln p_{\sigma}-\lambda \left(1-\frac{p_{-\sigma}}{p_{\sigma}}\right) - k \delta (x) .
\end{equation}
Integrating over $ x\in [-\epsilon, \epsilon]$ with $\epsilon \rightarrow 0$ determines the boundary condition at the target location, $x=0$, as
\begin{equation}
\ln \frac{p_{\sigma}(0^+)}{p_{\sigma}(0^{-})} = - \frac{\sigma k}{v},
\end{equation}
or equivalently,
\begin{equation}
\label{BC}
p_{\sigma}(0^+)=p_{\sigma}(0^{-}) \, e^{- \sigma k/v} .
\end{equation}
We note that the same boundary condition can also be derived using the Doi model~\cite{Whitehouse2013, Doi1976}:
When the particle is within the reaction radius from the target, e.g., $x\in[-\ell/2,\ell/2]$,
they react in the Poisson process with a constant rate $k/\ell$.
Then, letting $\ell \rightarrow 0$, one shows that the same boundary condition, Eq.~\eqref{BC}, is obtained.

The confining environments is specified by hard walls:
When RTPs reach a confining boundary, they get stuck until the next tumbling occurs to reverse their directions of motion.
This leads to accumulation of particles, developing delta-peaked distributions, at the walls.
Thus, a probability, not a probability density, for particles to be at $x=\pm L$ becomes finite~\cite{Angelani_2017}.
The continuity equation for the probability of the boundary layer can be obtained from Eq.~\eqref{master}:
Integrating Eq.~\eqref{master} over $[L-\epsilon, L]$ in the limit of $\epsilon \rightarrow 0$, we obtain
\begin{equation}
\partial_t w_{+} (L, t) = v p_+ (L-\epsilon, t) - \lambda w_{+} (L, t)
\end{equation}
and 
\begin{equation}
0 = -v p_- (L-\epsilon, t) + \lambda w_{+} (L, t)
\end{equation}
where the probability of the boundary layer is defined as
\begin{equation}
w_{\sigma} (L, t) \equiv \lim_{\epsilon \rightarrow 0} \int_{L-\epsilon}^{L} \, dx \, p_\sigma (x, t) .
\end{equation}
Here, we used that $v p_\sigma (L, t) = 0$ as there is no particle current across the hard-wall,
and $w_- (L, t) = 0$ as the left-oriented particles do not accumulate at the right wall.
Repeating the same procedure, i.e., integrating over $[-L, -L+\epsilon]$, the continuity equation at the left wall can be obtained.
Combining them, the equations for $w_\sigma$ read 
\begin{equation}
\label{w_continuity}
\partial_t w_{\sigma} (t) = v p_\sigma (\sigma L, t) - \lambda w_\sigma (t)
\end{equation}
and the boundary condition for $p_{-\sigma}$ at $x=\sigma L$ is given as
\begin{equation}
 v p_{-\sigma}(\sigma L, t)= \lambda w_{\sigma} (t),
\end{equation}
where $w_\sigma (t) = w_\sigma (\sigma L, t)$ and from now on, it is understood that $p_\sigma (\sigma L, t) = \lim_{\epsilon \rightarrow 0} p_\sigma (\sigma (L- \epsilon), t)$.
Physical interpretation of Eq.~\eqref{w_continuity} is obvious. 
It is a conservation equation for the boundary layer probability $w_\sigma$ where the probability gain is given by the incoming flux, $v p_\sigma (\sigma L, t)$,
and the probabiliy loss is given by the fraction of particles flipping their directions at the wall~\cite{Angelani_2017}.

Now we rescale the space $X=x/L$ and time $T=\lambda t$.
Then the master equation becomes
\begin{equation}\label{normMaster0}
\partial_T P_\sigma=-\sigma\tilde{v}\partial_X P_{\sigma} - (P_\sigma-P_{-\sigma})
\end{equation}
for $0<|X|<1$, and the boundary conditions are given at the target location ($X=0$) as
\begin{equation}
P_\sigma(0^{+}) = P_{\sigma}(0^{-})e^{-\sigma \tilde{k}/\tilde{v}}\label{BC0}
\end{equation}
and at the hard-walls ($X=\sigma$) as
\begin{equation}
\partial_T w_\sigma=\tilde{v}P_{\sigma}(\sigma)-w_\sigma,  ~~~
0 = \tilde{v} P_{-\sigma}(\sigma )- w_\sigma\label{BCw}
\end{equation}
where $P_\sigma(X)$ is the probability distribution for the rescaled variable $X$, i.e. $P_{\sigma}(X) = Lp_\sigma(x)$. The rescaled velocity $\tilde{v}$ and the rescaled absorption strength (reactivity) $\tilde{k}$ are defined by 
\begin{equation}
\tilde{v}=\frac{v}{\lambda L}, ~~~
\tilde{k}=\frac{k}{\lambda L}.
\end{equation} 
If we define the survival probability as an integration of the probability distribution over the space $S(T)=\sum_\sigma \left[ \int_{-1}^{1}dX  P_\sigma(X) + w_\sigma(T) \right]$, the mean searching time can be obtain by integrating the survival probability over time:
\begin{eqnarray}
\langle T\rangle&=&-\int_0^\infty dT\, T \partial_T S(T) 
=\int_0^\infty  dT\,S(T) \nonumber \\
&=& \sum_{\sigma = \pm 1}\left[ \int_{-1}^{1} dX \phi_{\sigma}(X) + W_\sigma\right] .
\label{MFPT}
\end{eqnarray}
with the time-integrated probabilities $\phi_\sigma(X)$ and $W_\sigma$ defined by
\begin{eqnarray}
\phi_\sigma(X) &=& \int_0^{\infty} dT\, P_{\sigma}(X) , \\
 W_{\sigma} &=& \int_0^{\infty} dT\, w_\sigma.
 \end{eqnarray}
By integrating Eq.(\ref{normMaster0}) and Eqs.(\ref{BC0})-(\ref{BCw}) over time, we get ordinary differential equations for $\phi_\sigma (X)$:
\begin{equation}\label{backwardEq}
-\frac{1}{4}=-\sigma\tilde{v}\partial_X \phi_{\sigma} - (\phi_\sigma-\phi_{-\sigma})
\end{equation}
with boundary conditions
\begin{eqnarray}
\phi_\sigma(0^{+}) &=& \phi_{\sigma}(0^{-})e^{-\sigma \tilde{k}/\tilde{v}}\label{BC0int}\\
W_\sigma&=&\tilde{v}\phi_{\sigma}(\sigma)=\tilde{v} \phi_{-\sigma}(\sigma ).\label{BCwint}
\end{eqnarray}
Here, it is assumed that the initial probability distribution of the particle is uniform, i.e., $P_{\sigma}(X) = 1/4$.
Introducing 
$f(X) = (\phi_+ + \phi_{-})/2$ and $g(X) = (\phi_+ - \phi_{-})/2$,
Eq.(\ref{backwardEq}) can be written as
\begin{eqnarray}
-\frac{1}{4} &=& - \tilde{v} \partial_X g\label{deq1}\\
0 &=& -\tilde{v} \partial_X f - 2 g.\label{deq2}
\end{eqnarray}
Since $f(X)=f(-X)$ and $-g(X)=g(-X)$ by the symmetry, it suffices to consider only the right half-side, $X \in [0, 1]$.
The boundary conditions are given at the target position ($X=0^+$) as
\begin{equation}
\label{BC0_f}
\left(e^{\tilde{k}/\tilde{v}}-1\right)f(0^+)=-\left(e^{\tilde{k}/\tilde{v}}+1\right)g(0^+)
\end{equation}
and on the right wall ($X=1$) as
\begin{equation}
\label{BCg}
g(1)=0  \, , ~~~
W_\sigma=\tilde{v}f(1)\, .
\end{equation}
Integration of Eq.~\eqref{deq1} with Eq.~\eqref{BCg} gives
\begin{equation}
g(X) = -\mathrm{sgn}(X) \left( \frac{1-|X|}{4\tilde{v}} \right).
\end{equation}
Using $g(X)$ and Eq.~\eqref{BC0_f}, we integrate Eq.~\eqref{deq2} to find
\begin{equation}
f(X) = \frac{1}{e^{\tilde{k}/\tilde{v}}-1}\frac{1}{2\tilde{v}} +\frac{1}{4\tilde{v}}+ \frac{2|X|-X^2}{4\tilde{v}^2} ,
\end{equation}
which leads to
\begin{equation}
W_+=\frac{1}{2}\frac{1}{e^{\tilde{k}/\tilde{v}}-1} +\frac{1}{4}+ \frac{1}{4\tilde{v}} .
\end{equation}
Then, the mean searching time $\langle T\rangle$ is evaluated by using Eq.~\eqref{MFPT} as
\begin{equation}\label{MFPTD0}
\langle T\rangle = \frac{2}{3\tilde{v}^2}+\frac{3}{2\tilde{v}}+\frac{1}{2}+\frac{1}{e^{\tilde{k}/\tilde{v}}-1}\left(1+\frac{2}{\tilde{v}} \right) .
\end{equation}
This is one of the main results of our study, which gives an explicit expression for the mean searching time by a RTP
in a one-dimensional confined space in the presence of a partially absorbing target represented by a finite reactivity $k$.

\begin{figure}[tp]
\centering
\includegraphics[width=\linewidth]{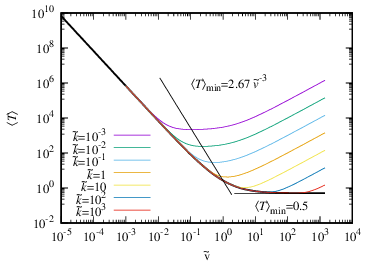}
\caption{The rescaled mean searching time $\langle T\rangle$ for a one-dimensional run-and-tumble searcher is plotted using the analytic expression, Eq.(\ref{MFPTD0}), as a function of rescaled velocity $\tilde{v}$ for various rescaled reaction rates $\tilde{k}$ of the target. 
The black thick line corresponds to a perfectly absorbing boundary ($\tilde{k}=\infty$), and thin solid lines with different colors represent $\langle T\rangle$ for different $\tilde{k}$ values as indicated in the legend.
The minimum value of mean searching time lies on the line $\langle T\rangle_{min}=8 / (3\tilde{v}^{3})$ for $\tilde{k}\ll 1$ and converges to $1/2$ for $\tilde{k}\gg 1$ (see the main text).
 }
\label{fig1-1}
\end{figure}

In Fig.\ref{fig1-1}, the mean searching time $\langle T \rangle$ is presented as a function of $\tilde{v}$ for various $\tilde{k}$, using the analytic expression of Eq.(\ref{MFPTD0}).
If the target is perfectly absorbing ($\tilde{k}\rightarrow \infty$), the mean searching time decreases monotonically as $\tilde{v}$ increases and converges to $1/2$ for $\tilde{v}\rightarrow \infty$.
On the other hand, if $\tilde{k}$ is finite, non-monotonic behavior of the mean searching time is observed, where $\langle T \rangle$ decreases for small values of $\tilde{v}$, while it increases linearly in $\tilde{v}$ at large $\tilde{v}$.
The optimal value $\tilde{v}^*$ that minimizes the mean searching time is an increasing function of $\tilde{k}$.

There are several studies on the target search by an active particle where a perfectly absorbing target is considered in higher dimensions, 
and the dependence of the searching time on $\tilde{v}$ presented in Fig.\ref{fig1-1} is qualitatively different from what these studies predict.
If the searcher exhibits the persistent random walk in two(or three)-dimensional lattice with periodic boundary conditions, the mean first passage time is a monotonically decreasing function of self-propulsion velocity when the flipping rate is fixed\cite{PhysRevLett.108.088103}.
The same is true for a RTP with a target located at the center of two(or three)-dimensional circular(or spherical) domain\cite{rupprecht2016optimal}.
If the searcher is an ABP with a translational diffusion, on the other hand, it has been observed that there is a parameter range that the mean first passage time becomes a non-monotonic function of self-propulsion speed when the target is at the center of two-dimensional circular domain\cite{doi:10.1063/1.4952423}.
In this case, however, the searching time increases exponentially as the self-propulsion velocity increases\cite{doi:10.1063/1.4952423}, unlike ours where the searching time grows linearly in $\tilde{v}$.

\section{Asymptotics of the mean searching time}\label{sec3}
\subsection{Decomposition into the first-passage time and the first-return time}
To obtain a comprehensive picture of the underlying physics, we express the searching time $T$ for a partially absorbing target in the following form:
\begin{equation}\label{Tdecomp}
T = T_{first}+ \sum_{i = 1}^{N_{pass}} T_{return}^{(i)} 
\end{equation}
where $T_{first}$ is {\it first-passage time}, i.e., the time it takes for a searcher starting from a given initial position to arrive at the target for the first time, 
$T_{return}^{(i)}$ is the {\it first-return time} to the target starting from the target position after $i$-th visit to the target,
and $N_{pass}$ is the number of times that the searcher passes the target until it is finally absorbed.
Since they are all independent random numbers, the average of searching time can be written as
\begin{equation}
\langle T\rangle = \langle T_{first}\rangle+ \langle N_{pass}\rangle \langle T_{return}\rangle  \label{MFPTgener}
\end{equation}
where $\langle T_{return}\rangle = \langle T_{return}^{(i)}\rangle$.
The boundary condition at $x=0$, Eq.~\eqref{BC}, determines the probability that RTP finds the target when visiting the target site,
$p_{find}=1-e^{-\tilde{k}/\tilde{v}}$.
The probability distribution of $N_{pass}$ is thus given as
\begin{equation}\label{pdfN}
P(N_{pass}) = (1-p_{find})^{N_{pass}}p_{find},
\end{equation}
which yields
\begin{equation}
\langle N_{pass}\rangle  = \frac{1-p_{find}}{p_{find}} = \frac{1}{e^{\tilde{k}/\tilde{v}} - 1} .
\end{equation}

$\langle T_{first}\rangle$ and $\langle T_{return}\rangle$ are the mean first-passage times for different initial distributions of searcher; 
$\langle T_{first}\rangle$ is the mean first-passage time when the initial searcher distribution is uniform, and $\langle T_{return}\rangle$ is the mean first-passage time when the searcher starts from the target site.
In Appendix A and B, the mean and the variance of $T_{first}$ and $T_{return}$ are analytically evaluated to give
\begin{eqnarray}
\langle T_{first}\rangle &=& \frac{2}{3\tilde{v}^2}+\frac{3}{2\tilde{v}}+\frac{1}{2}\label{Tfir} , \\
\langle T_{return}\rangle&=&1+\frac{2}{\tilde{v}} \label{Ti} .
\end{eqnarray}
By substituting these into Eq.(\ref{MFPTgener}), the expression of the mean searching time, Eq.(\ref{MFPTD0}), is reproduced.

In the following subsections, we investigate the asymptotic behaviors of $\langle T \rangle $ in different limits 
and find the optimal velocity that minimizes the searching time.

\subsection{Mean searching times in the diffusive limit and in the ballistic limit}
Let us begin with two limiting cases: $\tilde{v}\rightarrow 0$ limit and $\tilde{v}\rightarrow \infty$ limit.
The rescaled velocity $\tilde{v}$ is, by definition, the ratio between the persistence length $\xi=v/\lambda$ and the system size $L$.
Depending on whether $\tilde{v}$ is small or large, the motion of searcher is considered to be diffusive or ballistic.

In the limit of $\tilde{v}\rightarrow 0$, i.e., the persistence length $\xi$ is much smaller than the system size $L$,  the trajectory of a run-and-tumble particle resembles that of a diffusing particle with an effective diffusion coefficient $\tilde{D}_{eff}=\tilde{v}^2/2$. 
Then, $\langle T_{first}\rangle$ can be identified with the mean first-passage time of a diffusive searcher, 
\begin{equation}\label{diffusive}
\langle T_{first}\rangle = \frac{1}{3 \tilde{D}_{eff}} \simeq  \frac{2}{3\tilde{v}^2}.
\end{equation}
As $\tilde{v}$ decreases, the effective diffusion coefficient decreases, so the mean first-passage time increases.
In the same limit, $\langle N_{pass}\rangle\langle T_{return}\rangle$ is vanishingly small:
Although the mean returning time [Eq.(\ref{Ti})] increases as the velocity $\tilde{v}$ decreases, the average number of passing becomes exponentially smaller, $\langle N_{pass}\rangle\simeq e^{-\tilde{k}/\tilde{v}}$ for large $\tilde{k}/\tilde{v}$.
The net returning time $\langle N_{pass}\rangle\langle T_{return}\rangle$ is given as
\begin{equation}
\langle N_{pass}\rangle\langle T_{return}\rangle \simeq \frac{2}{\tilde{v}}e^{-\tilde{k}/\tilde{v}}
\end{equation}
which goes to zero as $\tilde{v}\rightarrow 0$.
In the limit of $\tilde{v} \rightarrow 0$, the mean searching time is dominated by $\langle T_{first}\rangle$,
\begin{equation}
\langle T\rangle \simeq \langle T_{first}\rangle \simeq  \frac{2}{3\tilde{v}^2}.
\end{equation}
which is a decreasing function of $\tilde{v}$.

In the opposite limit of $\tilde{v}\rightarrow \infty$, the searcher undergoes the ballistic motion to the confining boundaries.
When the searcher reaches the hard-wall boundary, it exerts pressure against the wall and stays stuck until the next tumbling occurs to reverse its orientation.
As the direction of the searcher flips 
the searcher leaves this boundary, passes the target almost instantaneously, and finds the target with a probability of $1-e^{-\tilde{k}/\tilde{v}}$.

In this limit, the particle spends most of its time stuck at the hard wall boundary, waiting for the flipping of the direction. 
As normalized by the inverse of the flipping rate, the mean returning time is given as
\begin{equation}
\langle T_{return}\rangle\simeq1 ,
\end{equation}
since a single flipping is needed for returning, which takes one on average, 
and the mean first-passage time is given as
\begin{equation}
\langle T_{first}\rangle\simeq \frac{1}{2},
\end{equation}
because the number of flips required for a particle to first reach its target is either 0 or 1 with the same probability, depending on its initial orientation.
Thus the total mean searching time is determined by the number of passing $\langle N_{pass}\rangle$, which diverges as $\tilde{v}$ increases.
In the limit of $\tilde{v}\rightarrow \infty$,  the total returning time $\langle N_{pass}\rangle\langle T_{return}\rangle$ dominates the first-passage time $\langle T_{first}\rangle$, and the mean searching time is well-approximated as
\begin{equation}\label{balistic}
\langle T\rangle \simeq \langle N_{pass}\rangle \langle T_{return}\rangle \simeq [e^{\tilde{k}/\tilde{v}}-1]^{-1}\simeq \frac{\tilde{v}}{\tilde{k}},
\end{equation}
which is an increasing function of $\tilde{v}$.

\subsection{Asymptotic behavior of the optimal velocity}
Since the searching time diverges for both $\tilde{v}\rightarrow 0$ and $\tilde{v}\rightarrow \infty$, there exists an optimal velocity $\tilde{v}^*$ that minimizes the searching time, as we observe in Fig.\ref{fig1-1}.
The optimal velocity is determined by the interplay between the first passage time $\langle T_{first}\rangle$ and the total returning time $\langle N_{pass}\rangle \langle T_{return}\rangle$.
Here we separate two cases, large $\tilde{k}$ and small $\tilde{k}$, and investigate how the optimal velocity $\tilde{v}^*$ depends on $\tilde{k}$ for each case.

Let us first suppose that $\tilde{k}\ll 1$.
In this case, we can show that the optimal velocity $\tilde{v}^*$ is in the range of $\tilde{k}\ll\tilde{v}^*\ll 1$:
the optimal velocity must be within the diffusive regime $\tilde{v}^*\ll 1$. Otherwise, the total returning time  $\langle N_{pass}\rangle \langle T_{return}\rangle \sim O(\tilde{v}/\tilde{k})$ dominates the mean first-passage time $\langle T_{first}\rangle$ which is at most $\sim O(1)$. 
In the diffusive regime ($v \ll 1$), $\langle T_{first}\rangle \sim O(1/\tilde{v}^2)$ [Eq.(\ref{diffusive})] is much greater than  $\langle T_{return}\rangle \sim O(1/\tilde{v})$ [Eq.(\ref{Ti})], and for  $\langle T_{first}\rangle$ and $\langle N_{pass}\rangle \langle T_{return}\rangle$ to be comparable with each other, $\langle N_{pass}\rangle$ must be much greater than 1, which leads $\tilde{v}^* \gg \tilde{k}$. 

When $\tilde{k}\ll\tilde{v}\ll 1$, the average number of passing can be approximated as
\begin{equation}
\langle N_{pass} \rangle \simeq \frac{\tilde{v}}{\tilde{k}} -\frac{1}{2} + O(\tilde{k}/\tilde{v}) 
\end{equation}
and by multiplying with the mean returning time given by Eq.(\ref{Ti}), we have 
\begin{equation}\label{smallk2nd}
\langle N_{pass} \rangle \langle T_{return} \rangle \simeq  \frac{2}{\tilde{k}}+\frac{\tilde{v}}{\tilde{k}} -\frac{1}{\tilde{v}} - \frac{1}{2} + O(\tilde{k} ).
\end{equation}
Among the terms in Eq.(\ref{smallk2nd}), only the first and the second terms are relevant since other terms are always negligible compared to $\langle T_{first}\rangle \simeq 2/3\tilde{v}^2$.
Therefore, we can write $\langle T\rangle$ in this range of $\tilde{v}$ as
\begin{equation}
\langle T\rangle\simeq \frac{2}{3\tilde{v}^2}+\frac{2+\tilde{v}}{\tilde{k}},
\end{equation}  
and the optimal value $\tilde{v}^*$ is given as
\begin{equation}
\tilde{v}^*=\left(\frac{4}{3}\right)^{1/3} \tilde{k}^{1/3}. \label{vOptDif}
\end{equation}
The minimum value of $\langle T\rangle$ is then evaluated to be 
\begin{equation}
\langle T\rangle \simeq \frac{2}{\tilde{k}}+\frac{6^{2/3}}{2 \tilde{k}^{2/3}}\simeq \frac{2}{\tilde{k}} ,
\end{equation} 
and the minimum value of mean searching time can be expressed in terms of the optimal value $\tilde{v}^*$,
\begin{equation}
\langle T\rangle \simeq \frac{8}{3\tilde{v}^{*3}}
\end{equation}
as presented in Fig.\ref{fig1-1}.

Now, let us consider the case where $\tilde{k} \gg 1$.
In this regime, the optimal velocity must be much greater than 1 because, otherwise, $\langle N_{pass}\rangle \langle T_{return}\rangle$, which is at most $\sim O(e^{-\tilde{k}/\tilde{v}}/\tilde{v})$, would always be negligible compared to $\langle T_{first}\rangle$, which is a decreasing polynomial function of $\tilde{v}$.
For $\tilde{v}\gg 1$, the mean first-passage time is given as
 \begin{equation}
 \langle T_{first}\rangle \simeq \frac{1}{2}  + \frac{3}{2\tilde{v}} + O(\tilde{v}^{-2}),
 \end{equation}
and the leading order of $\langle N_{pass}\rangle$ and $\langle T_{return}\rangle$ are given as
\begin{eqnarray}
\langle N_{pass}\rangle &=& (e^{\tilde{k}/\tilde{v}}-1)^{-1} \\
 \langle T_{return}\rangle &\simeq& 1+O(\tilde{v}^{-1}).
\end{eqnarray}
Collecting the relevant terms in $\langle T\rangle$, we obtain
\begin{equation}
\langle T\rangle \simeq  \frac{1}{2}  + \frac{3}{2\tilde{v}} + \frac{1}{e^{\tilde{k}/\tilde{v}}-1},
\end{equation}
from which the optimal value of the rescaled speed $\tilde{v}$ is calculated as
\begin{equation}
\tilde{v}^* \simeq \frac{\tilde{k}}{\log\left(\frac{\tilde{2k}}{3}\right)} .
\end{equation}
The minimum value of the mean searching time is given as,
\begin{equation}
\langle T\rangle \simeq \frac{1}{2}+\frac{3}{2\tilde{k}}\left[1+\log \left(\frac{2\tilde{k}}{3}\right)\right]\simeq \frac{1}{2},
\end{equation}
which is a half of the mean flipping time
and correctly captures the behavior of $\langle T \rangle$ for $\tilde{k} \gg 1$ in Fig.~\ref{fig1-1}.

\section{Discussion}\label{sec4}

\subsection{First-passage problem of a diffusive RTP in higher dimension}\label{correspondence}
Here, we consider the first-passage problem of a diffusive RTP, i.e., a RTP subject also to diffusion, with a small perfect target in a three-dimensional elongated domain.
Despite the relevance of the translational diffusion in case of a small target in high dimensions, an analytic solution of this problem is unavailable to date.
As a possible application of the current result, we show that the first-passage time of a 3D diffusive RTP can be qualitatively understood in terms of a 1D RTP with a reactive target.

\begin{figure*}[t]
\centering
\includegraphics[width=\linewidth]{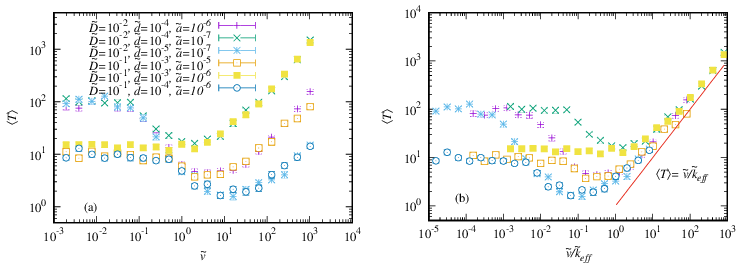}
\caption{(a) Simulation results of the rescaled mean first-passage time $\langle T\rangle$ as a function of rescaled velocity $\tilde{v}$, for a diffusive RTP in a three-dimensional box with two-dimensional array of perfectly absorbing spherical targets at $X=0$. The rescaled diffusion constant $\tilde{D}$, the spacing of the array $\tilde{d}$, and the radius $\tilde{a}$ of the target are presented in the figure. The mean first-passage times with the same $\tilde{D}$'s overlap for $\tilde{v}\ll 1$, while the mean first-passage times with the same $\tilde{k}_{eff}=4\pi \tilde{D} \tilde{a}/\tilde{d}^2$ are merged for $\tilde{v}\gg 1$.
(b) The rescaled mean first-passage time $\langle T\rangle$ as a function of $\tilde{v}/\tilde{k}_{eff}$ for the same parameters, indicating $\langle T \rangle \simeq \tilde{v}/\tilde{k}_{eff}$, as consistent with Eq.~\eqref{balistic}.
}
\label{fig2}
\end{figure*}

For this purpose, we consider a three-dimensional rectangular box, with hard wall boundary condition at $x=\pm L$, periodic boundary condition along the lateral directions, i.e., at $y$ (and $z$)$=\pm d/2$, and a spherical perfect target of a radius $a$ at the center ($\mathbf{r} =0$) which absorbs the searcher perfectly on the surface. 
We assume the high aspect ratio of the space and the small target size, $a\ll d \ll L$.
The motion of a diffusive RTP is  described by the overdamped Langevin equation,
\begin{equation}\label{master3d}
\frac{d}{dt}{\mathbf{r}} = v \, \mathbf{e}_{\Omega} + \sqrt{2D}\, {\boldsymbol \eta}(t)
\end{equation}
where $D$ is the diffusion constant, $\eta (t)$ a Gaussian white noise satisfying $\langle\eta_\mu(t)\eta_\nu(t^\prime)\rangle=\delta_{\mu\nu}\delta(t-t^\prime)$, and the tumbling occurs in the Poisson process with a rate $\lambda$, at which the self-propulsion direction $\mathbf{e}_\Omega$ is updated to a randomly chosen solid angle $\Omega$.
By performing the Langevin dynamics simulations of Eq.~\eqref{master3d}, we numerically estimate the mean first-passage times for various parameters, which are shown in Fig.~\ref{fig2}.
For numerical evaluations, we rescale all lengths by $L$ according to $\tilde{\mathbf{r}} = \mathbf{r}/L$ and introduce dimensionless velocity $\tilde{v} = v/ L \lambda$ and diffusion constant $\tilde{D} = D/L^2 \lambda$.
The first-passage time, measured in units of $\lambda^{-1}$ as $T = \lambda t$, is averaged over 100 realizations. 
The Fokker-Planck equation (FPE) for the probability distribution function of a diffusive RTP reads as
\begin{equation}
\label{FPE}
\partial_t p_{\Omega }=-v\mathbf{e}_\Omega\cdot \nabla_{\mathbf{r}} p_{\Omega}+{D}\nabla_{\mathbf{r}}^2 p_{\Omega}+\lambda \int\frac{d\Omega^\prime}{4\pi} \left[p_{\Omega^\prime}-p_{\Omega}\right]
\end{equation}
where $p_{\Omega }$ is the probability distribution function of the searcher to be at position $\mathbf{r}$ with self-propulsion direction $\Omega$.
$\ell_c = D/v$ is the length scale  separating a diffusion-dominant regime from a drift-dominant regime;
on a length scale of $\ell \ll \ell_c$, the diffusion term dominates in Eq.~\eqref{FPE}, while on a length scale of $\ell \gg \ell_c$, the drift term does.

When $L \ll D/v$ (or equivalently, $ \tilde{v} \ll \tilde{D}$), the diffusion is always dominant, i.e., even on the largest length scale of the system, $L$.
Trajectories of the particle then strongly resemble those of a diffusing particle and the first-passage dynamics can be well-approximated by that of a purely diffusive system.
For a 1D diffusive particle with a reactive target represented by a reaction rate $k_{eff}$, the mean first-passage time is given by
\begin{equation}
\langle t \rangle = \frac{L^2}{3 D_x} + \frac{L}{k_{eff}}
\end{equation}
or when rescaled using $\lambda^{-1}$,
\begin{equation}
\label{small_v}
\langle T \rangle 
= \frac{1}{\tilde{D}} + \frac{1}{\tilde{k}_{eff}}
\end{equation}
where the 1D diffusion constant $D_x$ is related to the 3D diffusion constant as $3 D_x = D$.
The 1D reaction rate, $k_{eff}$, can be estimated following the scheme described below [see Eq.~\eqref{keff}].
Our prediction, Eq.~\eqref{small_v}, is clearly supported by the numerical results of Langevin dynamics simulations, shown in Fig~\ref{fig2}(a):
For the regime of $\tilde{v} \ll \tilde{D}$, the mean first-passage time $\langle T \rangle$ does not depend on the propulsion speed $v$
and agrees well with Eq.~\eqref{small_v}.

Now we consider the opposite limit of $L \gg D/v$ (or equivalently, $ \tilde{v} \gg \tilde{D}$). 
In this case, depending on the length scales under consideration, either diffusion or drift prevails.
On the length scale of $\ell \sim L \gg \ell_c$, the diffusion term is negligible in Eq.~\eqref{FPE}, which leads to
\begin{equation}
\partial_t p_{\Omega }
\simeq  -v\mathbf{e}_\Omega\cdot \nabla_{\mathbf{r}} p_{\Omega}+\lambda \int\frac{d\Omega^\prime}{4\pi} \left[p_{\Omega^\prime}-p_{\Omega}\right] .
\end{equation}
This suggests that the self-propulsion velocity manifests itself in the first-passage dynamics, occurring over a length scale of $L$, 
and that the first-passage problem reduces to that of a RTP.
What remains in order to apply our results derived in the previous sections is to determine the 1D reaction rate, $k_{eff}$.
The reaction (absorption) rate is identified with the number of particles arriving at the target boundary per unit time, 
which is determined by the local concentration profile of the particle {\it close to the target}.
When we are interested in the vicinity of the target, i.e., on a length scale of $\ell \sim a \ll D/v$, 
the diffusion dominates so that the particle concentration profile near the target obeys the diffusion equation,
\begin{equation}
\partial_t {p} \simeq  D\nabla_\mathbf{r}^2 {p},
\end{equation}
where ${p}=\int d\Omega \, p_{\Omega}$. 
The 3D reaction rate $\kappa$ is then found from a solution of the diffusion equation by integrating the particle current over the target surface,
\begin{equation}
\kappa = \frac{1}{p_{\infty}}\int_{|\mathbf{r}|=a} d^{2}\mathbf{r}\, (-\hat{r}) \cdot \mathbf{J} = \frac{1}{p_{\infty}}\int_{|\mathbf{r}|=a} d^{2}\mathbf{r}\,D\partial_r p_{st},
\end{equation}
where $p_{st}$ is the steady state solution of the Laplace's equation $\nabla^2 p = 0$.
Considering the boundary conditions, $p(\mathbf{r})|_{r=a}=0$ and $p(\mathbf{r})|_{r\rightarrow \infty}=p_{\infty}$,
one recovers the well-known result of the reaction rate, $\kappa = 4\pi D a$, obtained for a spherical perfect target of radius $a$ in 3D with a spherical symmetry.
Now we relate this 3D reaction rate $\kappa$ to the 1D reaction rate $k_{eff}$.
For this, we adopt the scheme proposed in \cite{Grebenkov2022} in which a 2D reactive boundary layer at $x=0$, containing the target, was considered.
The main idea is to replace a small perfectly absorbing target in 3D with a 2D reactive layer which absorbs the particles through the cross-sectional area $S$ with an effective reaction rate $k_{eff}$~\cite{Grebenkov2022}.
At a steady state, the total number of particles absorbed through this reactive layer surface is equal to the number of particles arriving at the target surface,
which yields 
\begin{equation}\label{keff}
k_{eff} =  \frac{\kappa}{S} \simeq \frac{4\pi D a}{d^2} ,
\end{equation} 
where $S \simeq d^2$ for the 2D planar layer considered here.
Assuming that except the small region around the target, the concentration profile of particles is almost constant and can be considered homogeneous along the lateral directions, 
the projection is carried out onto the $x$-axis,
and the 2D reactive layer is reduced to the 1D reactive boundary with the reaction rate $k_{eff}$.

Using this $k_{eff}$, we conjecture that 
the first-passage problem of a 3D diffusive RTP can be captured by means of a RTP in 1D with a reactive target.
To illustrate this point, the mean first-passage time $\langle T \rangle$ is plotted in Fig.~\ref{fig2} as a function of self-propulsion velocity for various parameters.
When $\tilde{v}\gg \tilde{D}$,  $\langle T \rangle$ exhibit a non-monotonic behavior in $\tilde{v}$, similarly to Fig.\ref{fig1-1}.
For $\tilde{v}\gg 1$, the mean first-passage times with different parameters merge with one another if the estimated $\tilde{k}_{eff}=4\pi \tilde{D} \tilde{a}/\tilde{d}^2$ are the same [Fig.~\ref{fig2}(a)].
To emphasize this, we plot $\langle T \rangle$ as a function of $\tilde{v}/\tilde{k}_{eff}$ in Fig.~\ref{fig2}(b) where
all the simulation results collapse on a single curve, which clearly indicates $\langle T \rangle \simeq \tilde{v}/\tilde{k}_{eff}$ for large $\tilde{v}$, as consistent with Eq.~\eqref{balistic}.
This supports our conjecture that the first-passage dynamics of active particles in elongated high dimensional systems can be qualitatively understood by using
the result presented in this work, i.e., the 1D RTP in the presence of a reactive target with an effective reaction rate $k_{eff}$.
This mapping is no longer valid when $v$ increases further to $a \sim D/v$ because in that case, the particle concentration around the target does not satisfy the diffusion equation and thus, $k_{eff}$ cannot be estimated as described above.

\subsection{Filtering active particles}\label{exper}
The existence of the optimal speed ${v}^*$ implies that we can in principle extract RTPs having speed $v^*$.
Consider a mixture of non-interacting RTPs with different self-propulsion speeds spread out in a one-dimensional confined space with a reactive target at the center.
It is then expected that the particles with velocities close to the optimal speed $v^*$ are more likely to be absorbed faster.

\begin{figure}[tp]
\centering
\includegraphics[width=\linewidth]{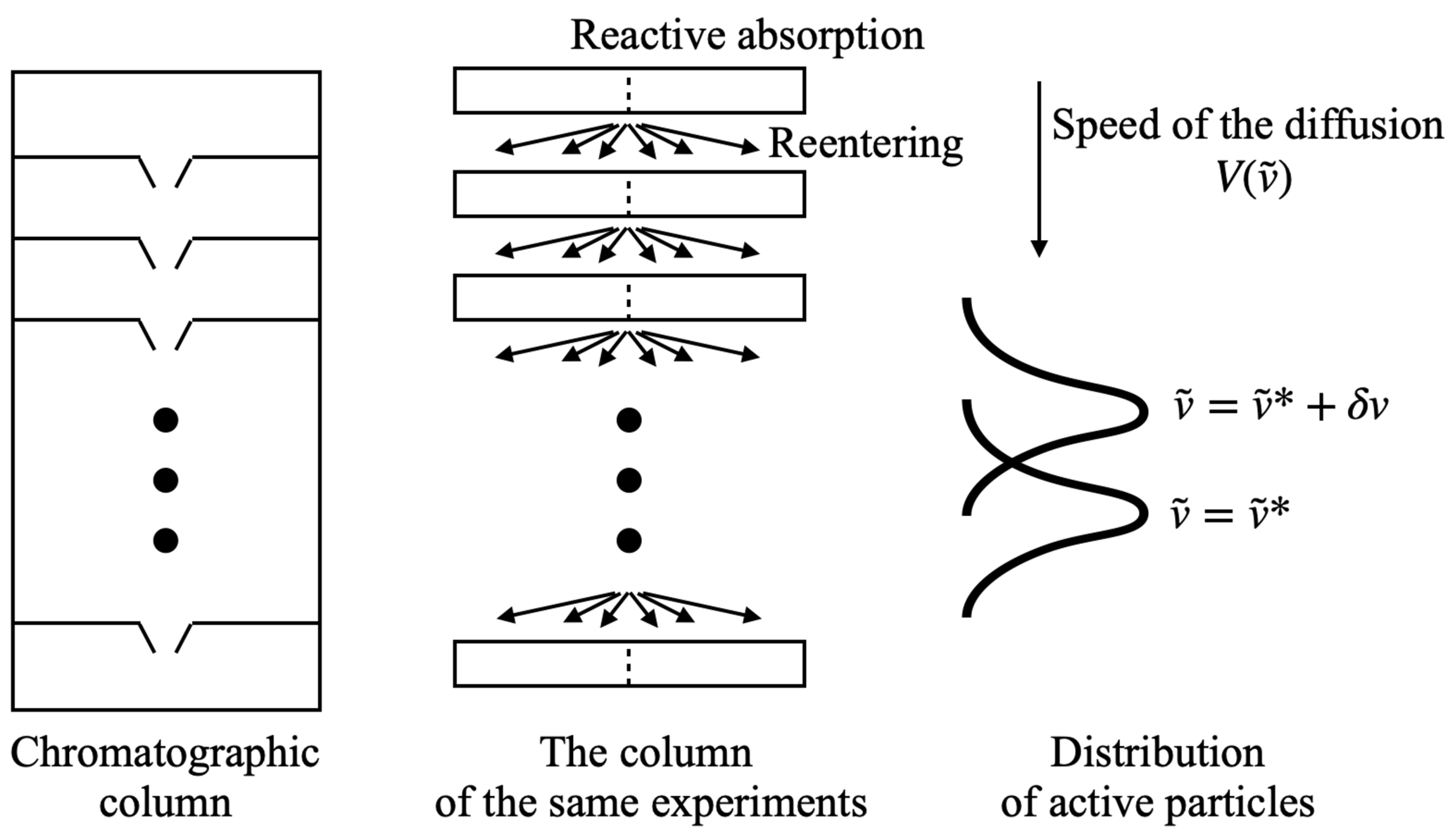}
\caption{Schematic figure for the active particle chromatography}
\label{fig3}
\end{figure}

In practice, this one-time attempt may be insufficient to achieve a meaningful resolution in filtering.
More efficient procedure can be conceptually designed as following. 
Many copies of the identical experimental set-up are prepared as a column, and a mixture of RTPs to be sorted is placed in the first set-up (see Fig.\ref{fig3}).
As soon as particles are extracted from one experiment, they are put into the next.
Over time, the particles of the first experimental set-up propagate into next set-ups, and the propagation speed depends on the mean searching time of each active particle.
Then we can finally obtain a chromatographic spectrum of active particles along the column, and the particle with optimal speed $\tilde{v}^*$ will be placed at the edge of the spectrum.

To be more specific, we can define the propagation speed along the column as the number of passed floors (set-ups) $N$ divided by the time it takes to migrate $T(N)$.
If $N$ is large, we can estimate $T(N)$ as
\begin{equation}
T(N)=N\langle T\rangle+\sqrt{N}\sigma_T \eta
\end{equation}
where $\sigma_T$ is the standard deviation of the mean first-passage time, which is evaluated in Appendix C [see, e.g., Eq.~(\ref{variance})], and $\eta$ is a standard normal random number. Then, the propagation speed $V$ along the column is written as
\begin{equation}
V=\frac{N}{N\langle T\rangle+\sqrt{N}\sigma_T \eta}\simeq \frac{1}{\langle T\rangle}+\sigma_V\eta^\prime
\end{equation}
and the standard deviation of the propagation speed is given as
\begin{equation}
\sigma_V=\frac{\sigma_T}{\sqrt{N}\langle T\rangle^2}.
\end{equation}
On the other hand, if the velocity $\tilde{v}$ of an active particle differs from $\tilde{v}^*$ by $\delta \tilde{v}$, the mean value of the propagation velocity differs from the optimal one by 
\begin{equation}
\delta V=\frac{\partial^2_{\tilde{v}}\langle T\rangle}{\langle T\rangle^2}\delta \tilde{v}^2.
\end{equation}
Here we define the resolution of the filtering as the ratio between $\tilde{v}$ and the minimum difference $\delta \tilde{v}_{min}$ that causes the difference in $V$ that can be distinguished from the standard deviation $\sigma_V$.
Since the minimum distinguishable difference $\delta\tilde{v}_{min}$ is given as
\begin{equation}
\delta \tilde{v}_{min}=\langle T\rangle\sqrt{\frac{\sigma_V}{\partial_{\tilde{v}}^2\langle T\rangle}}=\sqrt{\frac{\sigma_T}{\partial_{\tilde{v}}^2 \langle T\rangle}}N^{-1/4},
\end{equation}
 the resolution of the filtering is given as
\begin{equation}
R=\frac{\delta \tilde{v}_{min}}{\tilde{v}}=\frac{1}{\tilde{v}}\sqrt{\frac{\sigma_T}{\partial_{\tilde{v}}^2 \langle T\rangle}}N^{-1/4}.
\end{equation}
If $\tilde{k}\ll 1$, we can prove  $\partial_{\tilde{v}}^2 \langle T\rangle = (9/2)^{2/3}\tilde{k}^{-4/3}$ 
and $\sigma_T \simeq 2\tilde{k}^{-1}$,  and the resolution is given as
\begin{equation}
R=2^{1/6}3^{-1/3}\tilde{k}^{-1/6}N^{-1/4}
\end{equation}
and the height of the column must be at least $N\simeq 1.92v^{-2}R^{-4}$
 to achieve the resolution $R$.
If $\tilde{k}\gg1$, we have $\partial_{\tilde{v}}^2 \langle T\rangle \simeq \frac{3}{2} \tilde{k}^{-3} [\log\tilde{k}]^4$ and $\sigma_T=\sqrt{3/4}$, which gives
\begin{equation}
R=3^{-1/4}( \tilde{k}^{1/2}\log\tilde{k})N^{-1/4},
\end{equation}
which means that if we want to extract particles of self-propulsion speed $\tilde{v}\gg 1$ with a required resolution $R$, the height of the column, i.e., the minimum number of experimental setups must be at least $N\simeq 0.33\tilde{v}^{2}(\log\tilde{v})^4R^{-4} $.

\section{Summary}\label{sec5}
In this work, we have studied the random target search by a run-and-tumble particle in one-dimensional confined space with a partially absorbing target.
We considered the target represented by a sink term in the master equation with a single parameter, the absorption strength or the reaction rate.
Then, the mean searching time is analytically obtained, which is a non-monotonic function of the speed of the particle when the flipping rate is fixed.
This implies the existence of an optimal speed of the particle that minimizes the mean searching time.

To understand the non-monotonic dependence of the mean searching time on speed $\tilde{v}$, we examined the asymptotic behaviors of the mean searching time in the diffusive limit and the ballistic limit.
In the diffusive limit ($\tilde{v}\rightarrow 0$), the mean searching time is dominated by the average time it takes for the searcher to pass the target for the first time, which is proportional to $\tilde{v}^{-2}$.
In the ballistic limit ($\tilde{v}\rightarrow \infty$), on the other hand, the mean searching time becomes linear in $\tilde{v}$ since the average number that the searcher passes the target grows linearly in $\tilde{v}$ while the average time interval between passings is finite. 
The asymptotic behaviors of the optimal speed and the optimal searching time are investigated in the limit of large and small reaction rates.

We also showed that our result can provide a qualitative understanding of the first-passage problem of a diffusive run-and-tumble particle with a small perfectly absorbing target in a three-dimensional elongated domain. 
Finally, we proposed a conceptual design of filtering active particles with a certain speed. 
Since there is an optimal speed that minimizes the searching time, active particles with this optimal speed can be extracted from a mixture of active particles with different velocities by repeating quasi one-dimensional target searching experiments.
Here we provided an estimation of the minimum number of experiments required to achieve a specific resolution in filtration.

\acknowledgements
This research was supported by the National Research Foundation of Korea (NRF) grant funded by the Korean government (MSIT) (RS-2023-00251561).

\appendix

\section{Mean and variance of the first-passage time}
In this section, the mean and variance of the first-passage time for 1D run-and-tumble particle is evaluated by considering a perfect target at the origin.
Integrating Eq.(\ref{normMaster0}) over time, we obtain the differential equation for time-integrated probability
\begin{equation}\label{D0reverse}
-\frac{1}{4}=-\tilde{v}\sigma\partial_X \phi_{\sigma}-(\phi_\sigma-\phi_{-\sigma})
\end{equation}
for $0<|X|<1$.
At $X = \pm 1$, we get the boundary conditions by integrating Eqs.(\ref{BCw}) over time:
\begin{equation}\label{phiReflec}
\tilde{v}\phi_{\sigma}(X=\sigma)=\tilde{v}\phi_{-\sigma}(X=\sigma )=W_{\sigma},
\end{equation}
and at $X=0$, we have the perfect target described by a fully absorbing boundary condition:
\begin{equation}\label{phiReactive2}
\phi_\sigma(X=0^\sigma) = 0.
\end{equation}
Defining $f=(\phi_++\phi_-)/2$ and $g=(\phi_+ -\phi_-)/2$, equation (\ref{D0reverse}) becomes
\begin{eqnarray}
-\frac{1}{4}&=&-\tilde{v}\partial_Xg\\
0&=&-\tilde{v}\partial_X f-2 g,
\end{eqnarray}
and the boundary conditions [Eqs.(\ref{phiReflec}) and (\ref{phiReactive2})]  are written as
\begin{eqnarray}
0&=&f(X=0^{+})+g(X=0^{+})\\
0&=&g(X= 1)\\
W_{+}&=&\tilde{v} f(X=1).
\end{eqnarray}
Here, we only need to consider the domain satisfying $X>0$ and the remaining part is determined the symmetry $f(X)=f(-X)$, $g(X)=-g(-X)$, and $W_+ = W_-$.
As detailed in Sec. II, we solve the differential equations to find
\begin{eqnarray}
f&=& \frac{2X-X^2}{4\tilde{v}^2}+\frac{1}{4\tilde{v}}\label{phifir} ,\\
g&=&-\frac{1-X}{4\tilde{v}} ,\\
W_{+}&=&\frac{1}{4\tilde{v}}+\frac{1}{4},\label{Wfir}
\end{eqnarray}
for $X>0$, and the mean first-passage time can be obtained by 
\begin{eqnarray}
\langle T_{first}\rangle&=&2\int_{-1}^{1}dX \,f(X)+\sum_{\sigma=\pm} W_\sigma\\
&=&\frac{2}{3\tilde{v}^2}+\frac{3}{2\tilde{v}}+\frac{1}{2}.\label{mfrt}
\end{eqnarray}

The second order moment of a first-passage time $T$ (here, $T$ can be either $T_{first}$ or $T_{return}$) can be written in terms of the integration of the survival probability $S(T)$:
\begin{eqnarray}
\langle T^2\rangle&=&\int_0^{\infty}dT\, T^2 \left[-\frac{dS}{dT}\right]\\
 &=&2\int_0^{\infty}dT\, \left[-\frac{dS}{dT}\right] \int_0^T dT_1 \int_0^{T_1} dT_2 \\
&=&2\int_0^{\infty} dT_2 \int_{T_2}^\infty dT_1\int_{T_1}^{\infty}dT\, \left[-\frac{dS}{dT}\right]  \\
&=&2\int_0^{\infty}dT_2\int_{T_2}^{\infty} dT_1 S(T_1)\\
&=&2\sum_{\sigma=\pm}\left[\int_{-1}^{1} dX\, \Phi_\sigma(X) + \mathcal{W}_\sigma \right].
\end{eqnarray}
where $\Phi_\sigma=\int_0^{\infty}dT\int_T^{\infty} dT^\prime P_\sigma(T^\prime)$ and $\mathcal{W}_{\pm}=\int_0^\infty dT\int_T^\infty dT^\prime w_\pm(T^\prime)$.

To obtain $\langle T_{first}^2\rangle$, we derive the differential equation for $\Phi_\sigma$ by  integrating Eq.(\ref{normMaster0}) twice,
\begin{eqnarray}\label{EqPhi}
-\phi_\sigma&=&-\sigma\tilde{v}\partial_X \Phi_\sigma-(\Phi_\sigma-\Phi_{-\sigma}).
\end{eqnarray}
where $\phi_\sigma$ is the time-integrated probability that we already obtained. 
The boundary conditions at $X=\pm 1$ can be obtained by integrating Eqs.(\ref{BCw}) over time twice:
\begin{eqnarray}
-W_{\sigma}&=&\tilde{v}\Phi_\sigma (X=\sigma)-\mathcal{W}_{\sigma}\label{BCvar1}\\
0&=&\tilde{v}\Phi_{-\sigma} (X=\sigma)-\mathcal{W}_{\sigma},
\end{eqnarray}
and for a perfectly absorbing target at $X=0$, we have
\begin{equation}
\Phi_{\sigma}(X=0^\sigma) = 0.\label{BCvar3}
\end{equation}

Solving this using the expression of $\phi_\sigma = f + \sigma g$ and $W_{\pm}$ that we have obtained in Eqs.(\ref{phifir})-(\ref{Wfir}), we finally get
\begin{eqnarray}
F&=& \frac{1}{8\tilde{v}}+\frac{3+X^{2}}{8\tilde{v}^{2}}+\frac{2+9|X|-3X^{2}}{12\tilde{v}^{3}}\nonumber\\
&&-\frac{|X|^{3}-\frac{1}{4}X^{4}-2|X|}{6\tilde{v}^{4}}\\
G&=&\mathrm{sgn}(X)\left[\frac{3X^{2}-|X|^{3}-2}{12\tilde{v}^{3}}-\frac{3-2|X|}{8\tilde{v}^{2}}-\frac{1}{8\tilde{v}}\right]\nonumber\\ \\
\mathcal{W}_{+}&=&\frac{1}{4}+\frac{5}{8\tilde{v}}+\frac{2}{3\tilde{v}^{2}}+\frac{5}{24\tilde{v}^{3}} ,
\end{eqnarray}
where $\Phi_+ = F+G$ and $\Phi_- =F-G$.
Then, the second order moment and variance of $T_{first}$ are written as
\begin{eqnarray}
\left\langle T_{first}^{2}\right\rangle &=&2\sum_{\sigma=\pm}\left[\int_{-1}^{1}dX\,\Phi_{\sigma}(X)+\mathcal{W}_{\sigma}\right]\\
&=&1+\frac{7}{2\tilde{v}}+\frac{6}{\tilde{v}^{2}}+\frac{9}{2\tilde{v}^{3}}+\frac{16}{15\tilde{v}^{4}}\\
\left\langle T_{first}^{2}\right\rangle_c &=&\frac{3}{4}+\frac{2}{\tilde{v}}+\frac{37}{12\tilde{v}^{2}}+\frac{5}{2\tilde{v}^{3}}+\frac{28}{45\tilde{v}^{4}}.\label{varfrt}
\end{eqnarray}

\section{Mean and variance of the first return time}
The mean returning time is the mean first-passage time when the searcher starts its motion from the target location.
Therefore, the initial searcher distribution reads as
\begin{equation}
P_\sigma(X,T=0) = \frac{1}{2}\delta (X-0^\sigma),
\end{equation}
which accounts for that the particle is oriented to the right (or left) with probability of 1/2,
and the time integrated probability satisfies 
\begin{equation}
-\frac{1}{2} \delta (X-0^\sigma) = -\sigma \tilde{v} \partial_X \phi_\sigma - (\phi_\sigma -\phi_{-\sigma}).
\end{equation}
Integrating this over $X\in [0,\epsilon]$ with $0^+<\epsilon\ll 1$, we get a boundary condition for time-integrated probability at $X=\epsilon$,
\begin{eqnarray}\label{BCmrt}
\phi_+(\epsilon)  &=& \frac{1}{2\tilde{v}} 
 \end{eqnarray}
and in the remaining domain $X\in[\epsilon,1]$, the differential equation is written as
\begin{equation}
0=-\sigma \tilde{v} \partial_X \phi_\sigma - (\phi_\sigma -\phi_{-\sigma}).
\end{equation}
Solving this differential equation with boundary condition Eqs.(\ref{phiReflec}) and (\ref{BCmrt}), we obtain
\begin{eqnarray}
\phi_\pm&=&\frac{1}{2\tilde{v}}\label{phiret}\\
W_\pm&=&\frac{1}{2}\label{Wret}
\end{eqnarray}
and the mean returning time is found to be
\begin{equation}
\langle T_{return}\rangle = 1 + \frac{2}{\tilde{v}}.\label{mrt}
\end{equation}

The variance of the returning time can be evaluated from the double-time-integrated probability $\Phi_\sigma=\int_0^\infty dT\int_{T}^\infty dT^\prime P_\sigma (X,T^\prime)$, which satisfies Eq.(\ref{EqPhi}) with boundary conditions, Eqs. (\ref{BCvar1})-(\ref{BCvar3}).
Substituting Eqs.(\ref{phiret})-(\ref{Wret}) and solving the differential equation, we obtain 
\begin{eqnarray}
F&=&\frac{1}{4\tilde{v}}+\frac{1+|X|}{2\tilde{v}^{2}}+\frac{2|X|-X^{2}}{2\tilde{v}^{3}}\\
G&=&\mathrm{sgn}(X)\left[-\frac{1}{4\tilde{v}}-\frac{1-|X|}{2\tilde{v}^{2}}\right]\\
\mathcal{W}_{+}&=&\frac{1}{2}+\frac{1}{\tilde{v}}+\frac{1}{2\tilde{v}^{2}},
\end{eqnarray}
where $\Phi_+ = F+G$ and $\Phi_- =F-G$.
The second order moment and variance of $T_{i}$ are evaluated as
\begin{eqnarray}
\left\langle T_{return}^{2}\right\rangle &=&2+\frac{6}{\tilde{v}}+\frac{8}{\tilde{v}^{2}}+\frac{8}{3\tilde{v}^{3}}\\
\left\langle T_{return}^{2}\right\rangle_c &=&1+\frac{2}{\tilde{v}}+\frac{4}{\tilde{v}^{2}}+\frac{8}{3\tilde{v}^{3}}. \label{varrt}
\end{eqnarray}

\section{Mean and variance of the searching time}\label{chvariance}

Substituting Eq.(\ref{mfrt}) and Eq.(\ref{mrt}) into Eq. (\ref{MFPTgener}), the mean searching time is simply obtained as Eq.(\ref{MFPTD0}).
To evaluate the variance of the searching time, we start from the square of Eq.(\ref{Tdecomp});
\begin{equation}
 T^2 = \left(T_{first}+\sum_{i=1}^{N_{pass}}T^{(i)}_{return}\right)^2.
\end{equation}
Taking average over $T_{first}$ and $T^{(i)}_{return}$'s with a fixed value of $N_{pass}$, we have
\begin{eqnarray}
\left<T^2\right>_{N_{pass}}= \left(\left<T_{first}\right>+N_{pass}\left<T_{return}\right>\right)^2 \nonumber\\+ \langle T_{first}^2\rangle_c +N_{pass}\left<T_{return}^2\right>_c
\end{eqnarray}
and by taking average over $N_{pass}$, we get
\begin{eqnarray}
\left<T^2\right>&=& \left(\left<T_{first}\right>+\left<N_{pass}\right>\left<T_{return}\right>\right)^2 \nonumber\\
&&+ \langle T_{first}^2\rangle_c +\left<N_{pass}\right>\left<T_{return}^2\right>_c\nonumber\\
&& +\left<N_{pass}^2\right>_c \left<T_{return}\right>^2.
\end{eqnarray}
Thus, the variance of the first passage time is given as
\begin{equation}
\langle T^2\rangle_c = \langle T_{first}^2\rangle_c + \langle N_{pass}\rangle \langle T_{return}^2\rangle_c+\langle N_{pass}^2\rangle_c\langle T_{return}\rangle^2.
\end{equation}
From the probability distribution of $N_{pass}$ [Eq.(\ref{pdfN})], we obtain $\langle N_{pass}\rangle =[e^{\tilde{k}/\tilde{v}}-1]^{-1} $, $\langle N_{pass}^2\rangle_c =e^{\tilde{k}/\tilde{v}}[e^{\tilde{k}/\tilde{v}}-1]^{-2} $ and by substituting Eqs.(\ref{mrt}), (\ref{varfrt}) and (\ref{varrt}), the variance of the total searching time is evaluated as
\begin{eqnarray}
\langle T^2\rangle_c &=& \frac{3}{4}+\frac{2}{\tilde{v}}+\frac{37}{12\tilde{v}^{2}}+\frac{5}{2\tilde{v}^{3}}+\frac{28}{45\tilde{v}^{4}}
\nonumber\\
&&  + \frac{1}{e^{\tilde{k}/\tilde{v}}-1}\left[1+\frac{2}{\tilde{v}}+\frac{4}{\tilde{v}^{2}}+\frac{8}{3\tilde{v}^{3}}\right]\nonumber\\
&&+\frac{e^{\tilde{k}/\tilde{v}}}{(e^{\tilde{k}/\tilde{v}}-1)^2}\left(1+\frac{2}{\tilde{v}}\right)^{2}.\label{variance}
\end{eqnarray}
The standard deviation $\sigma_T$ of MFPT can be evaluated by taking square root of Eq.(\ref{variance}).



%

\end{document}